\documentclass[twocolumn,journal]{IEEEtran}
\usepackage[T1]{fontenc}
\usepackage[latin9]{inputenc}
\usepackage{float}
\usepackage{amstext}
\usepackage{amsmath}
\usepackage{flushend}
\usepackage{enumerate}
\usepackage{xcolor}

\usepackage{array}
\usepackage{multirow}

\usepackage{graphicx}

\usepackage{epstopdf}
\usepackage{amssymb}

\usepackage{amsthm,amsmath,amssymb}

\usepackage{mathrsfs}

\newtheorem{remark}{Remark}

\usepackage{subfigure}

\usepackage{graphics}
\usepackage{diagbox}

\usepackage{makecell}

\usepackage[ruled,vlined]{algorithm2e}
\usepackage{cite}

\usepackage[unicode=true,
 bookmarks=true,bookmarksnumbered=true,bookmarksopen=true,bookmarksopenlevel=1,
 breaklinks=false,pdfborder={0 0 0},pdfborderstyle={},backref=false,colorlinks=false]
 {hyperref}
\hypersetup{pdftitle={Your Title},
 pdfauthor={Your Name},
 pdfpagelayout=OneColumn, pdfnewwindow=true, pdfstartview=XYZ, plainpages=false}

\makeatletter

\floatstyle{ruled}
\newfloat{algorithm}{tbp}{loa}
\providecommand{\algorithmname}{Algorithm}
\floatname{algorithm}{\protect\algorithmname}

\makeatother

\begin{document}
\title{Sensing-Enhanced Handover Criterion for Low-Altitude Wireless Networks (LAWN)}
\author{Jingli~Li,~\IEEEmembership{Graduate Student Member,~IEEE,}~Yiyan~Ma,~\IEEEmembership{Member,~IEEE,}~Bo~Ai,~\IEEEmembership{Fellow,~IEEE,}\\
Weijie~Yuan,~\IEEEmembership{Senior Member,~IEEE,}~Qingqing~Cheng,~\IEEEmembership{Member,~IEEE,}~Guoyu~Ma,~\IEEEmembership{Member,~IEEE,}\\
Mi~Yang,~\IEEEmembership{Member,~IEEE,}~Yunlong~Lu,~\IEEEmembership{Member,~IEEE,}~Wenwei~Yue,~\IEEEmembership{Member,~IEEE,}\\
and~Zhangdui~Zhong,~\IEEEmembership{Fellow,~IEEE.}\\

\thanks{Jingli~Li, Yiyan~Ma, Bo~Ai, Guoyu~Ma, Mi~Yang, Yunlong~Lu, and Zhangdui~Zhong are with the School of Electronic and Information Engineering, Beijing Jiaotong University, Beijing 100044, China. Weijie~Yuan is with the School of System Design and Intelligent Manufacturing, Southern University of Science and Technology, Shenzhen 518055, China. Qingqing~Cheng is with the School of Electrical Engineering and Robotics, Queensland University of Technology, Brisbane, QLD 4000, Australia. Wenwei~Yue is with the State Key Laboratory of Integrated Services Networks, Xidian University, Xi\textquoteright an, Shaanxi 710071, China.}
}
\maketitle

\begin{abstract}

With the rapid growth of the low-altitude economy, the demand for cellular-enabled low-altitude wireless networks (LAWN) is rising significantly. The three-dimensional mobility of drones will lead to frequent handovers (HOs) in cellular networks, while traditional reference signal received power (RSRP)-based criteria may fail to capture the dynamic environment, causing redundant HOs or HO failures. To address this issue and motivated by the underutilization of sensing information in conventional HO mechanisms, we propose a novel HO activation criterion for drone systems that integrates both sensing parameters provided by integrated sensing and communication (ISAC) signals and RSRP. First, we construct an ISAC signal model tailored for low-altitude scenarios and derive the Cram\'er--Rao lower bound for sensing distance estimation. Subsequently, we propose a novel joint HO criterion that extends the conventional RSRP-based method by integrating sensing information from ISAC signals, enabling more reliable HOs in dynamic drone environments. Simulation results show that the joint HO criterion outperforms the baseline RSRP-based criterion under different signal-to-noise ratio (SNR) and sensing pilot ratio conditions. Particularly, when SNR \(\geq\) 0\,dB and the sensing pilot ratio is 20\%, the proposed joint HO criterion reduces the average HO region length by 75.20\% and improves the activation probability by 76.31\%.

\end{abstract}


\section{Introduction}

Emerging low-altitude applications are promoting the deployment of low-altitude wireless networks (LAWN) to support drone communications~\cite{yuan2025ground, jin2025co, liu2025movable}. Advanced cellular technologies, such as 5G-A and 6G, are expected to play a pivotal role in supporting these requirements~\cite{sun2024multi, sun2025dual, zhang2024generative, zhang2024g, zhang2024interactive}. Due to the relatively simplified radio propagation environment at low altitudes, the wireless channel between the drone and the base station (BS) is predominantly characterized by line-of-sight (LoS) propagation, creating favorable conditions for integrated sensing and communication (ISAC) applications~\cite{meng2024cooperative}. ISAC enables drones to simultaneously sense environmental parameters, such as delay and Doppler shifts from surrounding objects, while transmitting data~\cite{ma2024orthogonal}. Consequently, drones supported by ISAC can achieve enhanced environmental awareness while effectively utilizing sensing information to further improve communication reliability.

In LAWN, handover (HO) is a critical component in maintaining communication performance. The current 3GPP standard for 5G NR typically employs various events based on reference signal received power (RSRP) to trigger HO in traditional drone systems~\cite{3gpp.38.331}. However, in low-altitude scenarios involving high mobility and altitude variation, it struggles to cope with rapid channel fluctuations, resulting in redundant or failed HOs~\cite{jang2022uavs}. Compared to the traditional RSRP-based HO decision mechanism, ISAC introduces a spatial sensing dimension, providing a richer information basis for HO decisions. However, existing HO procedures remain largely tailored to traditional communication systems and have yet to fully utilize sensing capabilities offered by ISAC~\cite{3gpp.38.331}. Therefore, integrating ISAC sensing into the HO mechanism to enhance HO performance is a promising topic for LAWN.

To improve HO performance in LAWN, current studies have incorporated geometric or mobility information into RSRP-based HO mechanisms~\cite{jang2022uavs,hu2019trajectory}. However, such methods rely on external positioning systems, leading to extra hardware requirements and increased signaling overhead. Given the limitation, ISAC has gained attention for its ability to support concurrent sensing and communication within a unified framework. Towards that end, Meng \emph{et al}.~\cite{meng2024cooperative} developed a multi-BS cooperative localization mechanism to improve HO success rates by exploiting sensing information provided by ISAC, while Ge \emph{et al}.~\cite{ge2024target} designed a distributed HO based on trajectory sharing that enhances connection continuity, demonstrating the feasibility of applying ISAC to HO decision-making. Nevertheless, Lin \textit{et al.}~\cite{lin2023unified} advanced semantic communication and edge intelligence, offering insights into context-aware HO for LAWN. However, existing studies have not explored sensing-driven HO criteria or modeling the impact of sensing errors on HO behavior.

To bridge this gap, we propose a novel HO criterion that integrates sensing distance provided by ISAC signals and RSRP, with the main contributions summarized as follows:
\begin{itemize}
  \item We establish an ISAC signal model for LAWN and derive the Cram\'er--Rao lower bound (CRLB) for distance estimation. Based on the CRLB, we quantify the distance estimation error, serving as a foundation for designing a sensing-based HO criterion.
  \item A joint HO criterion is proposed by integrating the sensing distance parameter and RSRP, which triggers an HO whenever either the distance condition or the RSRP condition is met. By incorporating distance information, the proposed approach improves the HO success rate compared to RSRP-based methods.
  \item Simulation results show that, compared to the baseline RSRP-based criterion, the proposed joint HO criterion reduces the HO region length by 75.20\% and improves the HO activation probability by 76.31\% when signal-to-noise ratio (SNR) is greater than 0\,dB and the sensing pilot ratio is 20\%. Furthermore, it maintains performance improvements under low SNR and low sensing pilot ratio conditions.
\end{itemize}

The remainder of this paper is organized as follows: Section \ref{System Model} establishes the ISAC signal model and derives the CRLB for distance sensing. Section \ref{Handover} proposes the HO criterion. Section \ref{Performance Evaluation} provides performance simulations. Section \ref{Conclusion} concludes the paper.

\section{System Model}\label{System Model}

In this section, we establish the ground-to-air (G2A) geometric model, the ISAC signal model, and derive the CRLB for distance estimation.

\subsection{G2A Geometric Model}

\begin{figure}[t]
\centering \includegraphics[width=3.1in]{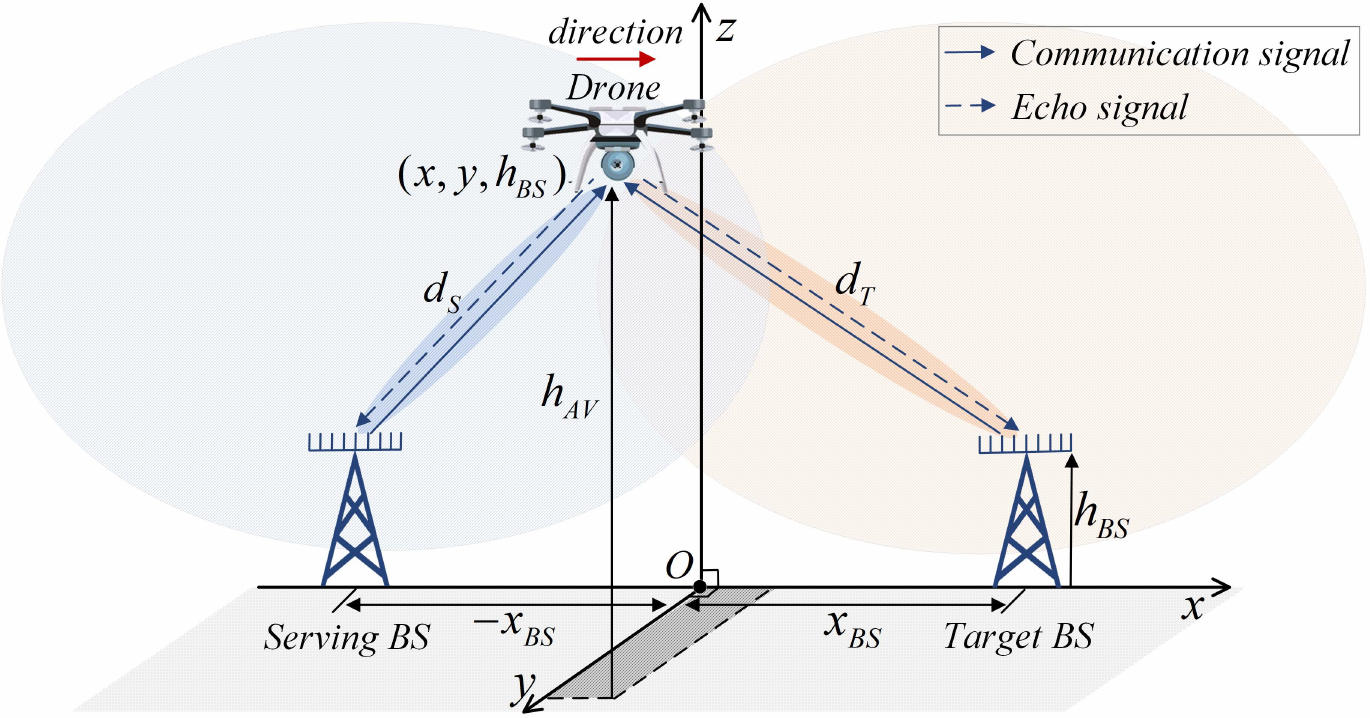}
\vspace{-1em}
\caption{G2A geometric model of the ISAC-enabled drone HO process.}
\vspace{-1em}
\label{shiyitu}
\end{figure}

In LAWN, the HO process of a drone from the serving BS to the target BS is shown in Fig.~\ref{shiyitu}. Without loss of generality, we establish a three-dimensional (3D) Cartesian coordinate system with the midpoint between the serving BS and the target BS as the origin, and the positive \(x\)-axis pointing toward the target BS. The antenna coordinates of the two BSs are \((-x_{\mathrm{BS}}, 0, h_{\mathrm{BS}})\) and \((x_{\mathrm{BS}}, 0, h_{\mathrm{BS}})\), respectively, while the drone is located at \((x, y, h_{\mathrm{AV}})\)~\footnote{Given the short duration of the HO process, the impact of drone velocity on HO decisions is negligible compared to that of distance. Therefore, as a preliminary study, this paper does not consider drone mobility during the HO decision-making procedure, which will be explored in future work.}. Let \( d_S \) and \( d_T \) denote the Euclidean distances from the drone to the serving BS and the target BS, respectively, then ${d_S} = \sqrt {{{(x + {x_{\mathrm{BS}}})}^2} + {y^2} + {{({h_{\mathrm{AV}}} - {h_{\mathrm{BS}}})}^2}} $ and ${d_T} = \sqrt {{{(x - {x_{\mathrm{BS}}})}^2} + {y^2} + {{({h_{\mathrm{AV}}} - {h_{\mathrm{BS}}})}^2}} $.

\subsection{ISAC Signal Model}

Assume that each BS is equipped with a uniform planar array (UPA) with ${N_t} = {N_x} {N_y}$ antenna elements, while the drone employs a single omnidirectional antenna. The BS transmits an orthogonal frequency-division multiplexing (OFDM) waveform that simultaneously enables communication and sensing, expressed as~\cite{ma2025channel}
\begin{equation}
s(t) = \sum\limits_{m = 0}^{M - 1} {\sum\limits_{n = 0}^{N - 1} {\sqrt {{P_\mathrm{sub}}} {a_{m,n}}{e^{j2\pi ({f_c} + n\Delta f)t}}{\rm rect}(\frac{{t - mT}}{T})} } ,
\label{eq:1}
\end{equation}
where \( M \) and \( N \) denote the number of OFDM symbols and subcarriers, respectively. The power allocated to the \( n \)-th subcarrier is \( P_{\mathrm{sub}} \), with uniform power allocation assumed~\cite{bansal2008optimal}, i.e., \( P_{\mathrm{sub}} = \frac{P_{\mathrm{sum}}}{N} \), where \( P_{\mathrm{sum}} \) represents the total transmit power. The modulation symbols \( a_{m,n} \) satisfy the normalization condition \( \mathbb{E}[|a_{m,n}|^2] = 1 \). We define the sensing pilot ratio~$\rho  \in (0,1]$ to represent the proportion of subcarriers used for sensing. Thus, given a total system bandwidth \( B \), the sensing bandwidth is ${B_S} = \rho B$, and the communication bandwidth is ${B_C} = \left( {1 - \rho } \right)B$. The carrier frequency is \( f_c \), and the subcarrier spacing is \( \Delta f = \frac{B}{N} \). The duration of each OFDM symbol is \( T = T_u + T_{\mathrm{CP}} \), where \( T_u = \frac{1}{\Delta f} \) is the useful symbol duration and \( T_{\mathrm{CP}} \) is the cyclic prefix duration. Besides, the function \( \rm{rect}(\cdot) \) in \eqref{eq:1} represents the windowing function.

For BS in full-duplex or time domain half-duplex mode, the signal transmitted by the BS is reflected by the target drone and received by the same BS for parameter estimation, and its time-domain expression is given by
\begin{equation}
\begin{array}{l}
\begin{aligned}
r(t) =& \sum\limits_{m = 0}^{M - 1} {\sum\limits_{n \in {\mathcal{S}_m}} {\sqrt {{P_\mathrm{sub}}} \mathbf{h}_{m,n}^H\mathbf{w}{a_{m,n}}{e^{j2\pi ({f_c} + n\Delta f)t}}} } \\
 &\times \operatorname{rect}(\frac{{t - {\tau_S} - mT}}{T}) + \omega(t) ,
\end{aligned}
\end{array}
\end{equation}
where $\mathcal{S}_m$ denotes the set of subcarrier indices allocated for sensing in the $m$-th OFDM symbol, with $\left| \mathcal{S}_m \right| = \rho N$. ${\tau _ S} = \frac{{2d}}{c}$ is the round-trip delay, \( c \) is the speed of light, and $\omega(t) \sim \mathcal{CN} \left( {0,{\sigma ^2}} \right)$ is the channel noise. $\mathbf{h}_{m,n}^H$ is the frequency-domain response of the sensing link at the \( n \)-th subcarrier and the \( m \)-th OFDM symbol, given by $\mathbf{h}_{m,n}^H = \sqrt {{\beta _S}} {e^{ - j2\pi n\Delta f{\tau _S}}}{\mathbf{v}^H}$, where $\sqrt {{\beta _S}}  = {10^{ - \frac{{{{\mathcal{L}}_S} + \varepsilon }}{{20}}}}$ is the sensing path gain, which accounts for the sensing path loss ${{\mathcal{L}}_S}$ and the shadow fading term $\varepsilon \sim \mathcal{N}(0, \sigma_{\text{SF}}^2)$. Here, ${{\cal L}_S} = 2{\cal L} - 10{\log _{10}}\sigma _{\rm RCS}  + 10\log \left( {\frac{{{\lambda ^2}}}{{4\pi }}} \right)$, where ${\cal L}$ denotes the communication path loss and ${\sigma _{\rm RCS}} $ is the radar cross section (RCS)~\cite{zhang2024cluster}. At typical drone flight altitudes of 100 to 300\,m, the LoS probability approaches 100\%~\cite{3gpp_tr_36_777}. Consequently, only the LoS path between the BS and drone is considered, with path loss is modeled using the 3GPP UMa-AV LoS model~\cite{3gpp_tr_36_777} as ${\cal L} = 28.0 + 22{\log _{10}}(d) + 20{\log _{10}}({f_c})$, where \( d \) is in meters and \( f_c \) is in GHz~\footnote{For distances below 100\,m, the probability of non-line-of-sight (NLoS) conditions increases, for which the UMa-AV NLoS model is adopted.}. Additionally, as a preliminary study, we assume perfect beam alignment for ISAC signal transmission, where the transmit beam direction \( \mathbf{w} \in \mathbb{C}^{N_x N_y \times 1} \) is aligned with the drone's array response vector \( \mathbf{v}^H \in \mathbb{C}^{1 \times N_x N_y} \), i.e., \( \mathbf{w} = \mathbf{v} \). In this case, the beamforming gain is $G = {\left| {{\mathbf{v}^H}\mathbf{w}} \right|^2} = N_t^2$.

By sampling the echo signal \( r(t) \), removing the cyclic prefix, and performing a fast Fourier transform (FFT), the resulting frequency-domain samples are given by
\begin{equation}
{r_{m,n}} = \sqrt {{P_{\mathrm{sub}}}{\beta _S}} {N_t}{a_{m,n}}{e^{ - j2\pi n\Delta f\tau _S }} + {\omega_{m,n}}, n \in {\mathcal{S}_m} .
\label{eq:r_mn}
\end{equation}

For the communication link, its frequency-domain channel model can be expressed as $\tilde{\mathbf{h}}_{m,n}^H = \sqrt {{\beta _C}} {e^{ - j2\pi n\Delta f{\tau _C}}}{\mathbf{v}^H}$, where ${\tau _C} = \frac{d}{c}$ is the one-way propagation delay, and $\sqrt {{\beta _C}}  = {10^{ - \frac{{{\cal L} + \varepsilon }}{{20}}}}$ is the large-scale gain. The received power at the drone is $P = {\beta _C}N_t^2{P_{\mathrm{sum}}}$, with its logarithmic form (dBm) as $P = {P_{\mathrm{sum}}} + 20{\log _{10}}{N_t} - {\cal L} - \varepsilon$. Additionally, the corresponding data rate \(R\) is given by~\cite{shannon1948mathematical}
\begin{equation}
R = ( 1- \rho) B{\log _2}\left( {1 + \frac{{{\beta _C}N_t^2{P_{\mathrm{sum}}}}}{{{N  \sigma ^2}}}} \right).
\label{eq:datarate}
\end{equation}

\subsection{CRLB Derivation}

Define the equivalent channel gain as \( \alpha \triangleq \sqrt{P_{\mathrm{sub}} \beta_S} N_t \), and the phase factor as \( A_n \triangleq 2\pi n \Delta f \). Assuming that the noise \( n_{m,n} \) is i.i.d. circularly symmetric complex Gaussian, the likelihood function of the observation vector \( \mathbf{r} \) parameterized by the time delay \( \tau_S \) is given by
\begin{equation}
L\left( {\mathbf{r}\left| \tau_S  \right.} \right) = \mathop \prod \limits_{m = 0}^{M - 1} \mathop \prod \limits_{n \in {\mathcal{S}_m}} \left[ {\frac{1}{{\sqrt {2\pi {\sigma ^2}} }}{e^{ - \frac{1}{{2{\sigma ^2}}}{{\left( {{r_{m,n}} - \alpha {a_{m,n}}{e^{ - j{A_n}\tau_S }}} \right)}^2}}}} \right].
  \label{eq:likelihood_function}
\end{equation}

Taking the logarithm of \eqref{eq:likelihood_function} and computing its second-order derivative with respect to \( \tau_S \), the Fisher information \( \mathbf{J}(\tau_S) \) is obtained as~\cite{mensing2013location}
\begin{equation}
\begin{array}{l}
\begin{aligned}
\mathbf{J}(\tau_S ) &=  - \mathbb{E}\left[ {\frac{{{\partial ^2}\ln \left( {L\left( {\mathbf{r}\left| \tau_S  \right.} \right)} \right)}}{{\partial {\tau_S ^2}}}} \right]\\
 & = \frac{\alpha }{{{\sigma ^2}}}\sum\limits_{m = 0}^{M - 1} {\sum\limits_{n \in {\mathcal{S}_m}} {A_n^2E\left[ {\operatorname{Re}\left\{ {{a_{m,n}}r_{m,n}^*{e^{ - j{A_n}\tau_S }}} \right\}} \right]} } .
\end{aligned}
\end{array}
\end{equation}

According to \( \mathbf{J}(\tau_S) \), the mean squared error of the time delay estimation satisfies \( \text{Var}(\hat{\tau_S}) \ge \frac{1}{\mathbf{J}(\tau_S)} \). Given the propagation model, the distance estimation is \( \hat{d} = \frac{c\hat{\tau_S}}{2} \), and its CRLB is \( \operatorname{CRLB}(d) = \frac{c^2}{4\mathbf{J}(\tau_S)} \). To relate estimation accuracy to the SNR, the average per-subcarrier SNR is defined as \( \gamma = \frac{\beta_S N_t^2 P_{\mathrm{sum}}}{N \sigma^2} \). Thus, \( \operatorname{CRLB}(d) \) can be rewritten as
\begin{equation}
\operatorname{CRLB}(d) = \frac{{3{c^2}}}{{8{\pi ^2}\gamma \Delta {f^2}M\rho N(\rho N - 1)(2\rho N - 1)}}.
\label{eq:CRLB}
\end{equation}
\begin{remark}
The theoretical lower bound $\operatorname{CRLB}(d)$ of distance estimation in \eqref{eq:CRLB} provides a theoretical foundation for the subsequent performance analysis of HO criteria. When $\rho N \ge 1$, considering \( \Delta f = \frac{B}{N} \) and ${B_S} = \rho B$, the lower bound can be approximated as $\operatorname{CRLB}(d) \approx \frac{{3{c^2}}}{{16{\pi ^2}\gamma \rho MNB_S^2}}$, indicating that sensing accuracy is inversely proportional to \( \gamma \), \( \rho \), \( M \), \( N \), and \( B_S^2 \), with particularly strong sensitivity to the sensing bandwidth \( B_S \).
\end{remark}

Assuming that the BS's distance estimation could achieve the CRLB and the estimation error follows a Gaussian distribution, i.e., \(\epsilon \sim \mathcal{N}(0, \operatorname{CRLB}(d))\), the estimated distance is modeled as \(\hat{d} \sim \mathcal{N}(d, \operatorname{CRLB}(d))\).

\section{HO Criterion and Performance Metrics}\label{Handover}

To enhance HO reliability, we propose a joint HO criterion that integrates the ISAC-based sensing information with the traditional RSRP-based mechanism. In the conventional RSRP-based HO criterion, the A3 event was adopted as the activation condition~\cite{3gpp.38.331}, where an HO is triggered when the received power \( P_T \) from the target BS exceeds that from the serving BS \( P_S \) by a hysteresis threshold \( \Gamma \), with the HO activation probability \( \mathbb{P}_{\mathrm{HO}}^{\mathrm{RSRP}} \) given by~\cite{hu2019trajectory}
\begin{equation}
\begin{array}{l}
\begin{aligned}
\mathbb{P}_{\mathrm{HO}}^{\mathrm{RSRP}} = \mathbb{P}\left[ {{P_T} - {P_S} > \Gamma } \right] = Q\left( {\frac{{\Gamma  + {\cal L}({d_T}) - {\cal L}({d_S})}}{{\sqrt 2 {\sigma _{SF}}}}} \right) .
\end{aligned}
\end{array}
\label{eq:HO1}
\end{equation}

In cellular networks, multipath fading and blockage hinder accurate user localization, prompting HO decisions to rely primarily on instantaneous communication measurements. In drone scenarios, however, the dominance of LoS links yields a more stable and observable spatial relationship between the drone and BSs, rendering distance a reliable basis for HO decisions. Based on this insight, we propose that HO can be triggered when the estimated distance \( \hat{d}_S \) to the serving BS exceeds that to the target BS \( \hat{d}_T \) by more than a threshold \( d_{\mathrm{th}} \), with the HO activation probability \( \mathbb{P}_{\mathrm{HO}}^{\mathrm{dist}} \) given by
\begin{equation}
\begin{array}{l}
\begin{aligned}
\mathbb{P}_{\rm{HO}}^{\rm{dist}} &= \mathbb{P}\left[ {{{\hat d}_S} - {{\hat d}_T} > {d_{\mathrm{th}}}} \right]\\
 &= Q\left( {\frac{{{d_{\mathrm{th}}} + {d_T} - {d_S}}}{{\sqrt {\operatorname{CRLB}({d_T}) + \operatorname{CRLB}({d_S})} }}} \right) .
\end{aligned}
\end{array}
\label{eq:HO2}
\end{equation}

To improve the robustness of HO decisions, we combine the sensing-based and RSRP-based criteria to form a joint HO mechanism, which triggers an HO when either condition is met, and the activation probability \( \mathbb{P}_{\mathrm{HO}}^{\mathrm{joint}} \) is given by
\begin{equation}
\begin{aligned}
\mathbb{P}_{\mathrm{HO}}^{\mathrm{joint}}
&= \mathbb{P} \left[ \left( P_T - P_S > \Gamma \right) \cup \left( \hat{d}_S - \hat{d}_T > d_{\mathrm{th}} \right) \right] \\
&= \mathbb{P}_{\mathrm{HO}}^{\mathrm{RSRP}}
+ \mathbb{P}_{\mathrm{HO}}^{\mathrm{dist}}
- \mathbb{P}_{\mathrm{HO}}^{\mathrm{RSRP}} \cdot \mathbb{P}_{\mathrm{HO}}^{\mathrm{dist}}~\footnotemark.
\end{aligned}
\label{eq:HO3}
\end{equation}
\footnotetext{The RSRP-based decision is primarily influenced by shadowing and channel noise in the communication link, while the error in sensing-based distance estimation in \eqref{eq:CRLB} mainly results from limited range resolution caused by constrained sensing bandwidth. Given the distinct physical origins and statistical independence between these two uncertainty sources, it is reasonable to assume independence between the RSRP-based and distance-based activation events as modeled in \eqref{eq:HO3}.}

To evaluate HO performance comprehensively, we introduce the effective data rate \( R_{\mathrm{eff}} \), which captures both activation probability and data rate, defined as
\begin{equation}
{R_{\mathrm{eff}}} = (1 - {\mathbb{P}_{\mathrm{HO}}}){R_S} + {\mathbb{P}_{\mathrm{HO}}}{R_T},
\end{equation}
where \( R_S \) and \( R_T \) denote the data rates when connected to the serving and target BSs, respectively. \( (1 - \mathbb{P}_{\mathrm{HO}}) R_S \) and \( \mathbb{P}_{\mathrm{HO}} R_T \) are the data rates before and after HO, respectively.

In terms of complexity, the proposed joint HO criterion requires additional distance information compared to traditional HO criteria. If distance is estimated using the joint parameter estimation method proposed in~\cite{xiao2024novel}, the resulting additional complexity is ${\cal O}\left[ \left( N_t + \log(MN) + g \right) \cdot NM \right]$, where $g$ denotes a resolution factor proportional to the size of the search unit. This indicates that the additional complexity remains comparable to that of conventional OFDM transceiver operations, ensuring a controllable computational overhead suitable for real-time online HO decisions in LAWN.

\section{Performance Evaluation}\label{Performance Evaluation}

\subsection{Simulation Parameter Settings}

To evaluate the performance of different HO criteria, we simulate the drone flying along the \( x \)-axis from the serving BS to the target BS at different altitudes as shown in Fig.~\ref{shiyitu}, with the simulation parameters listed in Table~\ref{simulation}~\footnote{To preliminarily evaluate the proposed HO scheme, a simplified drone trajectory is considered. More complex drone mobility patterns, such as 3D maneuvers and agile directional changes, will be explored in future work.}. Additionally, the sensing pilot ratio $\rho = 20\%$ is used as a nominal configuration to balance sensing and communication resources. According to \eqref{eq:HO1}--\eqref{eq:HO3}, the HO activation probability is shown to be a function of position, denoted as \( \mathbb{P}_{\mathrm{HO}} = f(x) \). We define the \( x \)-axis range corresponding to \( P_{\mathrm{HO}} \in [0.1, 0.9] \) as the HO region $D_{\mathrm{HO}} = \left[ {{f^{ - 1}}(0.1),{f^{ - 1}}(0.9)} \right]$, and its length is defined as the HO region length, given by $L_{\mathrm{HO}} = \left| f^{-1}(0.9) - f^{-1}(0.1) \right|$.

\begin{table}[!t]
\caption{The Main Simulation Parameters}\label{simulation}
\centering
\begin{tabular}{|l|l|}
  \hline
  \textbf{Parameters} & \textbf{Value}\tabularnewline
  \hline
  \hline
  Carrier frequency ${f_c}$ & 2\,GHz~\cite{3gpp.38.331} \\
  \hline
  System bandwidth $B$ & 10\,MHz~\cite{3gpp.38.331} \\
  \hline
  Transmission power of BS & 42\,dBm \\
  \hline
  BS antenna configuration ${N_x} \times {N_y}$ & 8 $\times$ 4~\cite{3gpp.38.331} \\
  \hline
  Coordinates of the serving BS antenna& (-1000, 0, 25)\,m \\
  \hline
  Coordinates of the target BS antenna& (1000, 0, 25)\,m \\
  \hline
  Drone altitude range &  (100, 300]\,m~\cite{3gpp.38.331} \\
  \hline
  Drone antenna configuration & Omni-directional \\
  \hline
  Number of OFDM symbols $M$ & 64~\cite{gaudio2019performance} \\
  \hline
  Number of subcarriers $N$ & 50~\cite{gaudio2019performance} \\
  \hline
  Subcarrier spacing $\Delta f$ & 200\,kHz \\
  \hline
  Duration of the effective symbol ${T_u}$ & 5\,$\mu s$ \\
  \hline
  Duration of the cyclic prefix ${T_{CP}}$ & 1.25\,$\mu s$ \\
  \hline
  Duration of the OFDM symbol $T$ & 6.25\,$\mu s$ \\
  \hline
  Speed of light $c$ & $3 \times {10^8}$\,m/s \\
  \hline
  Shadow fading $\sigma _{{\rm{SF}}}$ & $4.64\exp ( - 0.0066{\kern 1pt} {\kern 1pt} {h_{\mathrm{AV}}})$\,dB~\cite{3gpp.38.331} \\
  \hline
  RCS $\sigma _{\rm RCS}$ & 0.1\,${m^2}$~\cite{zhang2024cluster} \\
  \hline
  RSRP hysteresis $\Gamma$ & 2\,dB~\cite{3gpp.38.331} \\
  \hline
  Distance threshold ${d_{\mathrm{th}}}$ & 50\,m \\
  \hline
  Noise power ${{\sigma ^2}}$& -100\,dB \\
  \hline

\end{tabular}
\end{table}

\subsection{Simulation Analysis}

\subsubsection{Comparison of HO Activation Probability Distributions under Different HO Criteria}

\begin{figure}[!t]
    \centering
    \includegraphics[width=3.1in]{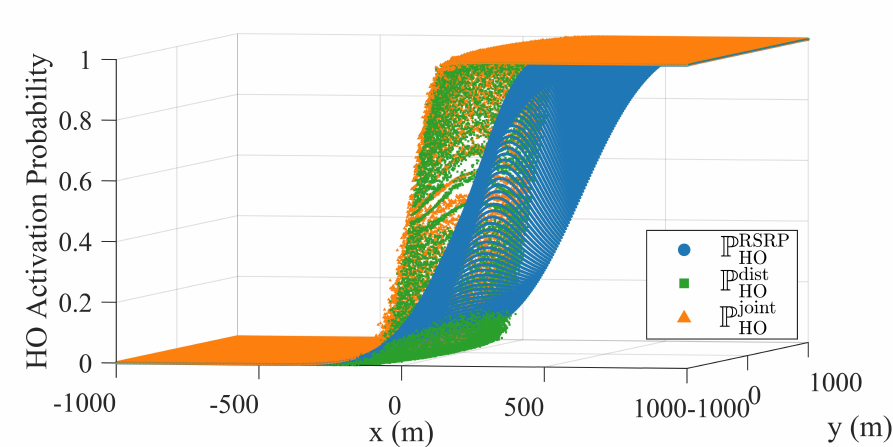}
    \vspace{-1em}
    \caption{HO activation probability distributions for various HO criteria at \( h_{\mathrm{AV}} = 200\,\text{m} \).}
    \label{tu1}
    \vspace{-1em}
\end{figure}

Fig.~\ref{tu1} shows the distribution of HO activation probabilities for RSRP-based, sensing-based, and joint HO criteria. First, under the RSRP-based HO criterion, the HO region exhibits the widest spatial distribution, with a statistical range of \( D_{\mathrm{HO}} = [-25, 465]\,\mathrm{m} \), corresponding to a length of 490\,m. Moreover, \( L_{\mathrm{HO}} \) increases from 235\,m to 490\,m as \( |y| \) increases. In contrast, the HO region under the sensing-based criterion is more concentrated, with a range of \( D_{\text{HO}} = [-65, 150]\,\mathrm{m} \), corresponding to a length of 215\,m. Moreover, the HO region length $L_{\mathrm{HO}}$ increases only moderately from 40\,m to 205\,m as $|y|$ increases. Finally, the joint HO criterion strikes a balance between the two, with a HO range of \( D_{\mathrm{HO}} = [-115, 130]\,\mathrm{m} \), corresponding to a length of 245\,m. Additionally, \( L_{\mathrm{HO}} \) increases from 60\,m to 220\,m as \( |y| \) increases. Overall, the joint HO criterion retains the early activation of the RSRP-based HO criterion while leveraging the spatial consistency and sensitivity of the sensing-based HO criterion, achieving a balanced trade-off between HO responsiveness and stability.

\subsubsection{Comparison of HO Activation Behavior under Different Thresholds}

\begin{figure}[!t]
    \centering
    \includegraphics[width=3.1in]{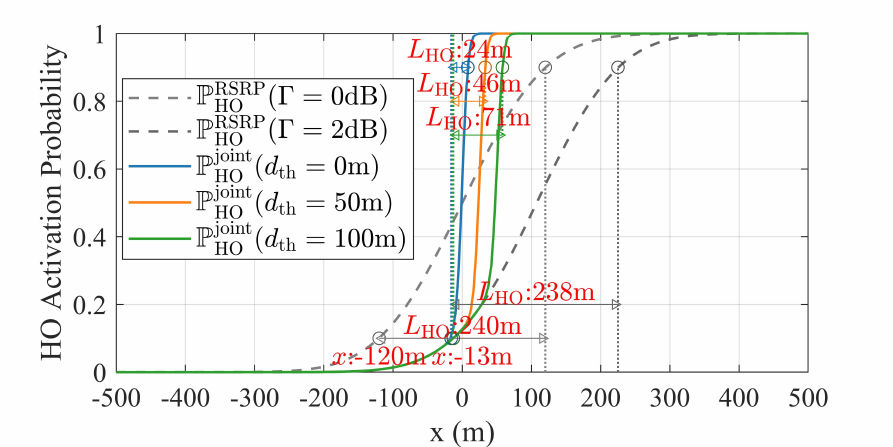}
    \vspace{-1em}
    \caption{HO activation probability under varying triggering thresholds for the RSRP-based and joint criteria, with \( y = 0 \) and \( h_{\mathrm{AV}} = 200\,\text{m} \).}
    \label{tu2}
    \vspace{-1em}
\end{figure}

Fig.~\ref{tu2} shows the HO activation probability of the RSRP-based and the joint HO criteria under different HO triggering thresholds. First, under the RSRP-based HO criterion, as \( \Gamma \) increases from 0 to 2\,\text{dB}, the lower boundary of the HO region $D_{\mathrm{HO}}$ shifts from \( x = -120\,\text{m} \) to \( x = -13\,\text{m} \), with a displacement of 107\,\text{m}. The corresponding \( L_{\mathrm{HO}} \) remains relatively large in all configurations. Second, under the joint HO criterion, despite \( d_{\mathrm{th}} \) spanning a wide range of 0--100\,\text{m}, the lower boundary of the $D_{\mathrm{HO}}$ remains consistently around \( x = -13\,\text{m} \), and \( L_{\mathrm{HO}} \) stays within a consistently small range across all configurations. We note that the joint HO criterion maintains a stable lower boundary and shorter HO region length under varying thresholds, demonstrating stronger robustness than the RSRP-based criterion. This robustness is attributed to the low-variance Gaussian modeling of sensing-based distance estimates, along with the complementary triggering mechanism of the joint HO criterion. In practical applications, this robustness can reduce the reliance on precise parameter tuning and enhance the predictability of HO behavior.

\subsubsection{Comparison of HO Activation Behavior under Different Sensing Pilot Ratios}\label{Pilot}

\begin{figure}[!t]
    \centering
    \includegraphics[width=3.1in]{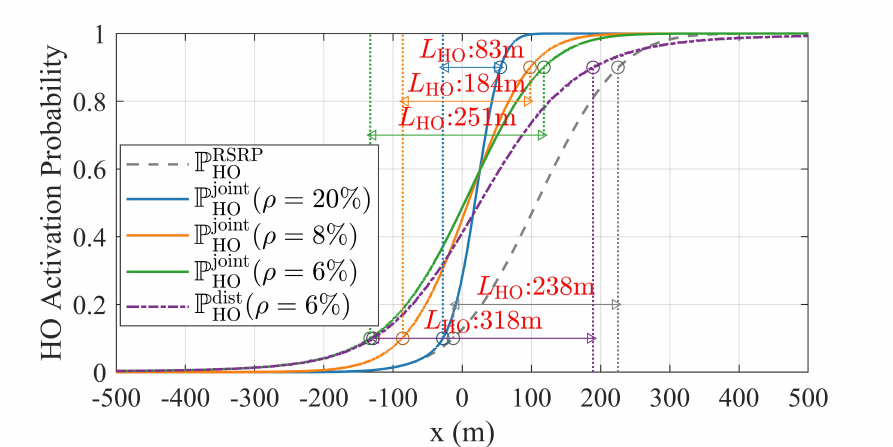}
    \vspace{-1em}
    \caption{HO activation probability under varying sensing pilot ratio under the joint HO criterion, with $ y = 0$ and $h_{\mathrm{AV}} = 200\,\text{m} $.}
    \label{tu3}
    \vspace{-1em}
\end{figure}

Fig.~\ref{tu3} shows the HO activation probability for the RSRP-based, sensing-based, and joint HO criteria under different sensing pilot ratios $\rho $. First, As $\rho$ increases, the \( L_{\mathrm{HO}} \) of the joint HO criterion exhibits a decreasing trend, indicating that the sensing distance parameter contributes to the improvement of HO performance. Second, when $\rho \geq 8\%$, the $L_{\mathrm{HO}}$ of joint HO criterion consistently remains lower than that of the traditional RSRP-based criterion, demonstrating superior performance in HO distance control. Finally, when $\rho$ decreases to 6\%, the $L_{\mathrm{HO}}$ of the sensing-based and joint HO criteria increase to 318\,m and 251\,m, respectively, both exceeding that of the RSRP-based HO criterion (238\,m). This is due to the increased estimation uncertainty caused by the decline in sensing accuracy. Under this extreme configuration, the joint HO criterion's $L_{\mathrm{HO}}$ is only slightly higher than that of the RSRP-based HO criterion. This indicates that under low sensing pilot ratio conditions, the joint HO criterion can rely on RSRP-based triggering to provide relatively stable HO decisions in the presence of sensing degradation, thereby suppressing excessive expansion of the HO region.

\subsubsection{Comparison of Average HO Region Lengths under Different Altitudes}

\begin{figure}[!t]
    \centering
    \includegraphics[width=3.1in]{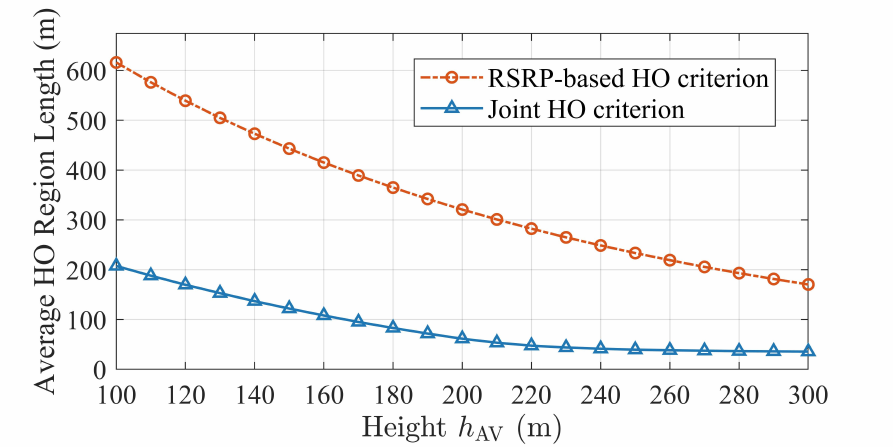}
    \vspace{-1em}
    \caption{Average HO region length versus altitude for the RSRP-based and joint HO criteria.}
    \label{tu5}
    \vspace{-1em}
\end{figure}

Fig.~\ref{tu5} shows the average HO region length \( L_{\mathrm{HO}} \) on the \( x\text{-}y \) plane under different altitudes \( h_{\mathrm{AV}} \), for both RSRP-based and joint HO criteria. As altitude increases, \( L_{\mathrm{HO}} \) of both HO criteria decreases: \( L_{\mathrm{HO}} \) of the RSRP-based HO criterion drops from 616\,m to 170\,m, while \( L_{\mathrm{HO}} \) of the joint HO criterion decreases from 208\,m to 36\,m. We note that the joint HO criterion consistently yields a smaller HO region length, with a 75.20\% average reduction in 3D space compared to the RSRP-based HO criterion.

\subsubsection{Comparison of Average HO Activation Probability Improvement under Different SNR conditions}

\begin{figure}[!t]
    \centering
    \includegraphics[width=3.1in]{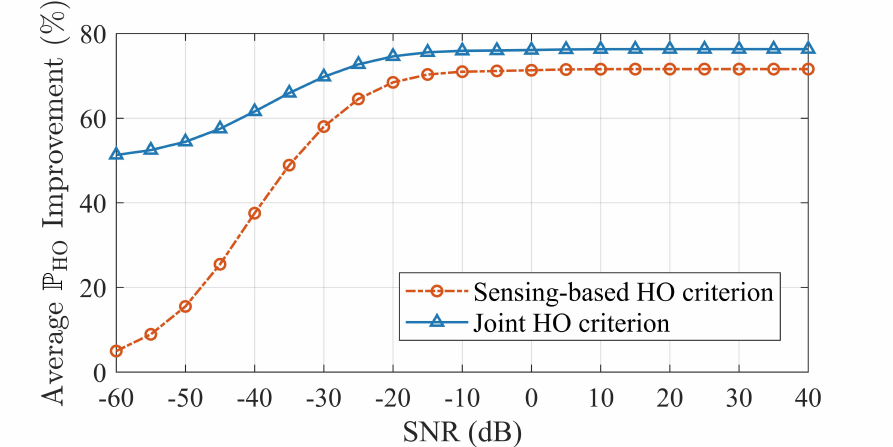}
    \vspace{-1em}
    \caption{Average improvement in HO activation probability versus SNR for the sensing-based and joint HO criteria.}
    \label{tu6}
    \vspace{-1em}
\end{figure}

Fig.~\ref{tu6} shows the average improvement in HO activation probability of the sensing-based and joint HO criteria over the RSRP-based criterion under various SNR within the 3D HO region. In the high SNR region (\( \text{SNR} \geq 0\,\text{dB} \)), the activation probability improvements of both HO criteria tend to stabilize, with the joint HO and sensing-based HO criteria reaching 76.31\% and 71.60\%, respectively. As the SNR decreases, the improvements of both criteria decline, and the performance gap gradually widens, while the joint HO criterion exhibits a smaller rate of decline. This can be explained by the theoretical lower bound: according to \eqref{eq:CRLB}, when \( \text{SNR} \leq -30\,\mathrm{dB} \), the CRLB rises to \( 10^2 \), making limited sensing accuracy a major constraint on the performance of the sensing-based HO criterion. In comparison, the joint HO criterion improves the HO activation probability by leveraging the sensing distance parameter, while the inclusion of the SNR-insensitive RSRP enables it to maintain a high activation probability even under degraded observation accuracy and exhibiting stronger noise robustness.

\subsubsection{Analysis of Data Rate Improvement}

\begin{figure}[!t]
    \centering
    \includegraphics[width=3.1in]{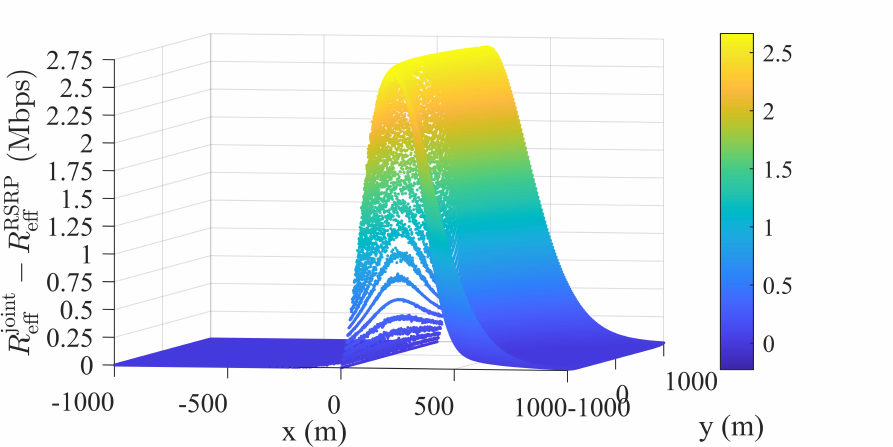}
    \vspace{-1em}
    \caption{Distribution of the effective data rate difference between the joint HO criterion and the RSRP-based HO criterion.}
    \label{tu4}
    \vspace{-1em}
\end{figure}

Beyond HO performance, the joint HO criterion also offers advantages in key service metrics, such as data rate. Fig.~\ref{tu4} shows the spatial distribution of the effective data rate difference \( R_{\mathrm{eff}}^{\mathrm{joint}} - R_{\mathrm{eff}}^{\mathrm{RSRP}} \) between the joint HO criterion and the RSRP-based HO criterion at \( h_{\mathrm{AV}} = 200\,\mathrm{m} \). The advantage of the joint HO criterion is primarily concentrated within the HO region, where the maximum data rate difference reaches 2.67\,Mbps; outside the HO region, the data rate of both criteria is nearly identical. This indicates that the joint HO criterion enhances transmission performance within the HO region, which is beneficial for drone applications demanding low latency and high reliability, such as real-time video and control signaling.

\section{Conclusion}\label{Conclusion}

To address the issue of redundant HOs or HO failures caused by the RSRP-based HO criterion in cellular-connected drone scenarios, we propose an HO criterion that integrates ISAC sensing information with RSRP to enhance HO performance. First, a low-altitude ISAC signal model is established, and the CRLB for distance estimation is derived. Based on this, we design a sensing-based HO criterion and integrate it with the RSRP-based HO criterion to form a joint HO criterion. Simulation results show that under conditions of SNR \(\geq 0 \, \text{dB}\) and a sensing pilot ratio of 20\%, the proposed joint HO criterion improves the handover activation probability by 76.31\% and reduces the HO region length by 75.20\%, significantly outperforming the baseline RSRP-based criterion. Moreover, even under challenging scenarios such as low SNR and reduced sensing pilot ratios, the proposed method maintains stable HO performance, demonstrating strong robustness.

This paper initially demonstrates the feasibility of sensing-assisted HO in cellular-connected LAWN. Future work will extend this study in several directions: First, realistic drone mobility models will be integrated into the HO decision process to better reflect low-altitude spatiotemporal characteristics. Second, trajectory-aware HO mechanisms will be explored to improve HO performance by utilizing drone flight path information. Third, the issue of ping-pong HO will be a key focus and will be addressed through enhanced HO mechanisms designed to reduce unnecessary HOs.

\bibliographystyle{ieeetr}
\bibliography{ref}

\end{document}